\documentstyle{EuroPhys}
\input EuroMacr
\begin{document} 

\addtolength{\evensidemargin}{0.50in}
\addtolength{\oddsidemargin}{0.50in}

\input epsf

\euro{40}{3}{281-286}{1997}
\Date{1 November 1997}
\shorttitle{Z. Toroczkai \etal 
BROWNIAN-VACANCY MEDIATED DISORDERING 
DYNAMICS}

\title{Brownian-Vacancy 
Mediated Disordering Dynamics}  
\author{Z. Toroczkai, G.
 Korniss\footnote{Present address:
Supercomputer Computations 
Research Institute, 
Florida State University,
Tallahassee Fl. 32306-4052, USA}, 
B. Schmittmann and R.K.P. Zia} 
\institute{
        Department of Physics and 
         Center for Stochastic 
         Processes in Science and Engineering, \\  
         Virginia Polytechnic 
          Institute and State University,\\  
         Blacksburg, VA. 24061-0435, USA}  
\rec{16 June 1997}{in final form 24 September 1997}

\pacs{
\Pacs{64}{60Cn}{Order-disorder 
transformations; statistical mechanics
of model systems}
\Pacs{68}{35Ct}{Interface 
structure and roughness}
\Pacs{05}{20$-$y}{Statistical mechanics}
     }

\maketitle 
 
\begin{abstract} 
The disordering of an initially 
phase segregated system of finite size, 
induced by the presence of highly 
mobile vacancies, is shown to exhibit 
dynamic scaling in its late stages.
 A set of characteristic exponents is 
introduced and computed analytically, 
in excellent agreement with Monte 
Carlo data. In particular, the 
characteristic time scale, controlling the 
crossover between increasing disorder
and saturation, is found to depend on 
the exponent scaling the number of  
vacancies in the sample. 
\end{abstract}

The structure and formation of 
surfaces and interfaces is of great 
scientific and technical importance
 and has attracted considerable interest 
over the past decade\cite{alb}. A 
second problem of similar relevance 
concerns the morphology and dynamics 
of ordering in bulk systems after a 
rapid temperature quench\cite{ord}. 
The evolution of spatial structures in 
both of these processes exhibits dynamic 
scaling in the late-time regime so 
that typical configurations at 
different times are self-similar after an 
appropriate rescaling of space 
and time. Clearly, the ``inverse'' scenario, 
i.e., the {\em destruction} of 
interfaces and the bulk {\em disordering} of 
an initially phase-segregated system,
 are also of major significance, being 
related to natural erosion phenomena 
\cite{er+corr}. Mobile defects, present 
in many materials, can obviously play
 a major role in the disordering 
process, especially if their dynamics 
is fast on the time scale governing 
the bulk particles. In this Letter, we
 consider a simple model for 
defect-mediated interface destruction 
and bulk disordering in a {\em finite} 
system, corresponding to a real 
material in which the characteristic time 
scale for vacancy diffusion is much 
faster than the ordinary bulk diffusion 
time. A fixed number of vacancies are 
initially located at a smooth, 
nonelastic interface separating two 
ordered bulk phases. These vacancies 
perform a homogeneous Brownian random 
walk, by exchanging positions with the 
otherwise completely passive bulk 
particles. As a result, the interface is 
gradually destroyed, and the system 
approaches a completely disordered 
equilibrium state. {\em Three} distinct
 temporal regimes, separated by two 
crossover times, are observed. Our key 
result is that the late stages of 
this process exhibit dynamic scaling.
 The characteristic exponents are 
computed analytically. In particular, 
the late crossover time is found to 
scale as a power of system size, 
determined by the spatial distribution of the 
defects. 
 
This Letter is organized as follows: 
We first describe our model, followed 
by a summary of our simulation results.
 Turning to analytic methods, we 
characterize the initial configuration
 and the final state exactly. The 
intermediate regime is described by a
 set of mean-field equations of motion 
which correctly predict both the scaling
 form for an appropriately defined 
``disorder parameter'' and the 
characteristic exponents controlling its size 
and time dependence. We conclude 
with some comments and questions. More 
details will be published elsewhere \cite{long}. 

\begin{figure}[htbp]
\vspace*{-0.6cm}
\hspace*{-0.5cm}
\begin{minipage}{1.55 in}  
\epsfxsize=1.55 in \epsfbox{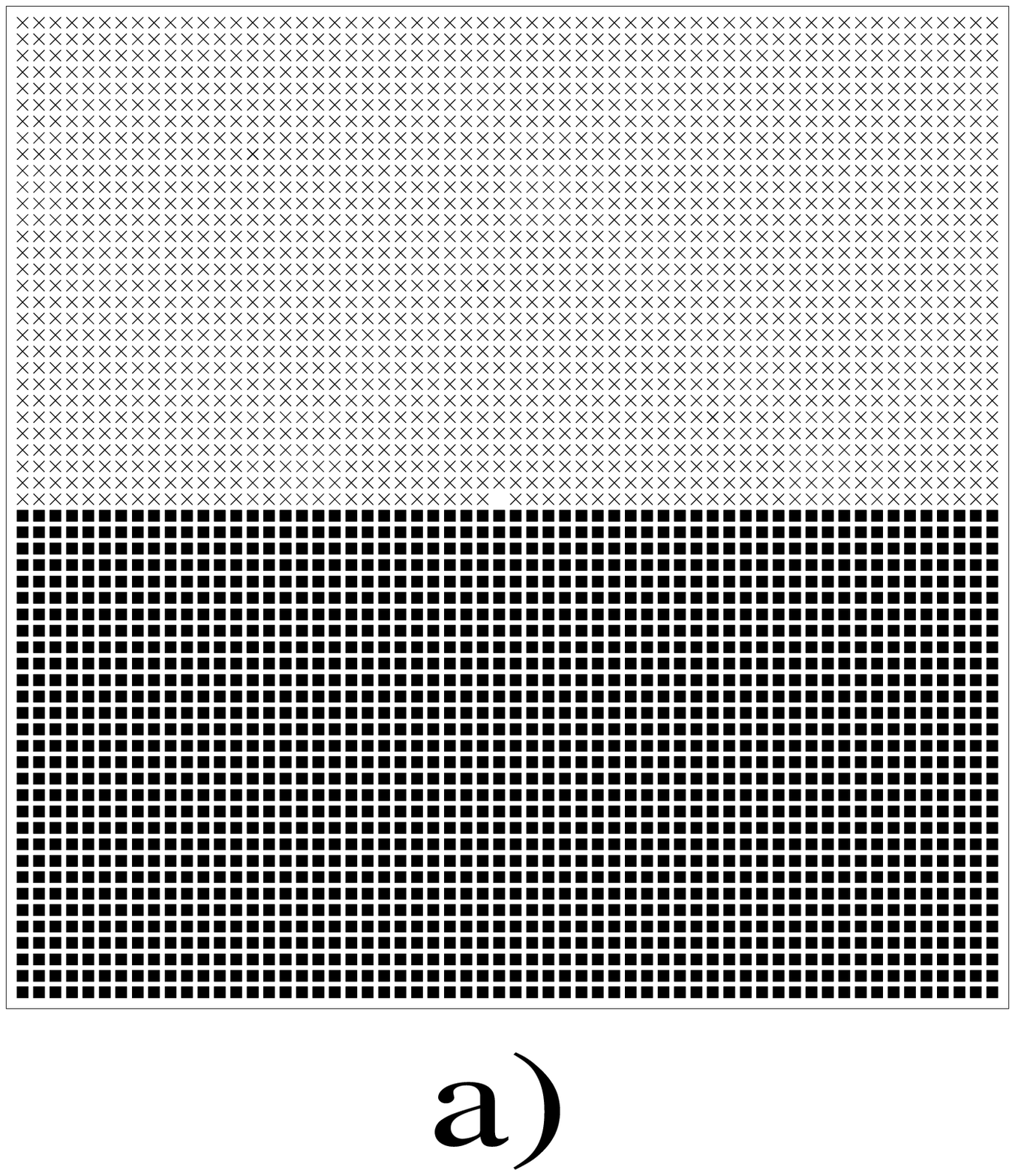}
\end{minipage}\hspace*{-0.5cm}
\begin{minipage}{1.55 in}  
\epsfxsize=1.55 in \epsfbox{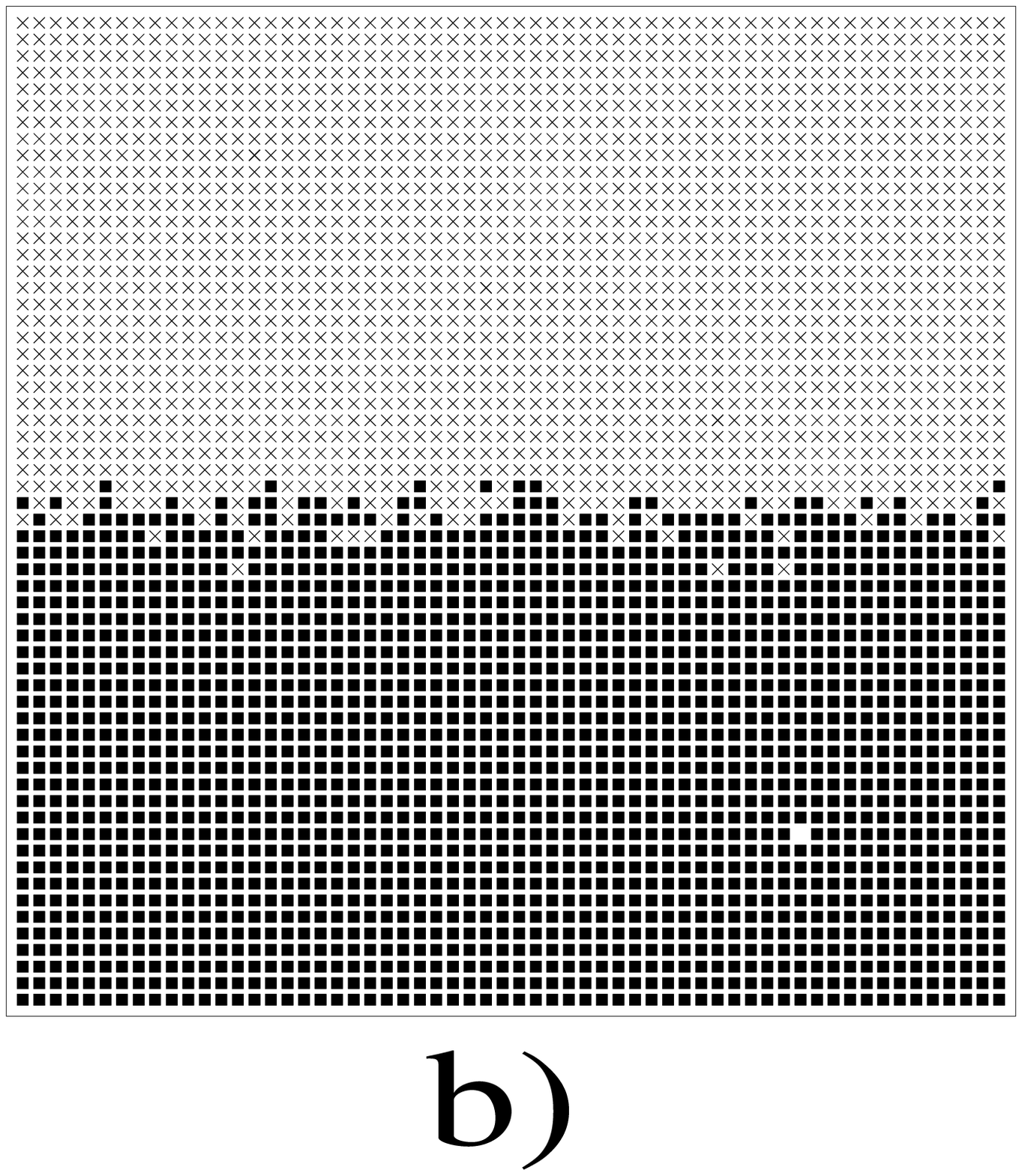}
\end{minipage} \hspace*{-0.5cm}                  
\begin{minipage}{1.55 in}  
\epsfxsize=1.55 in \epsfbox{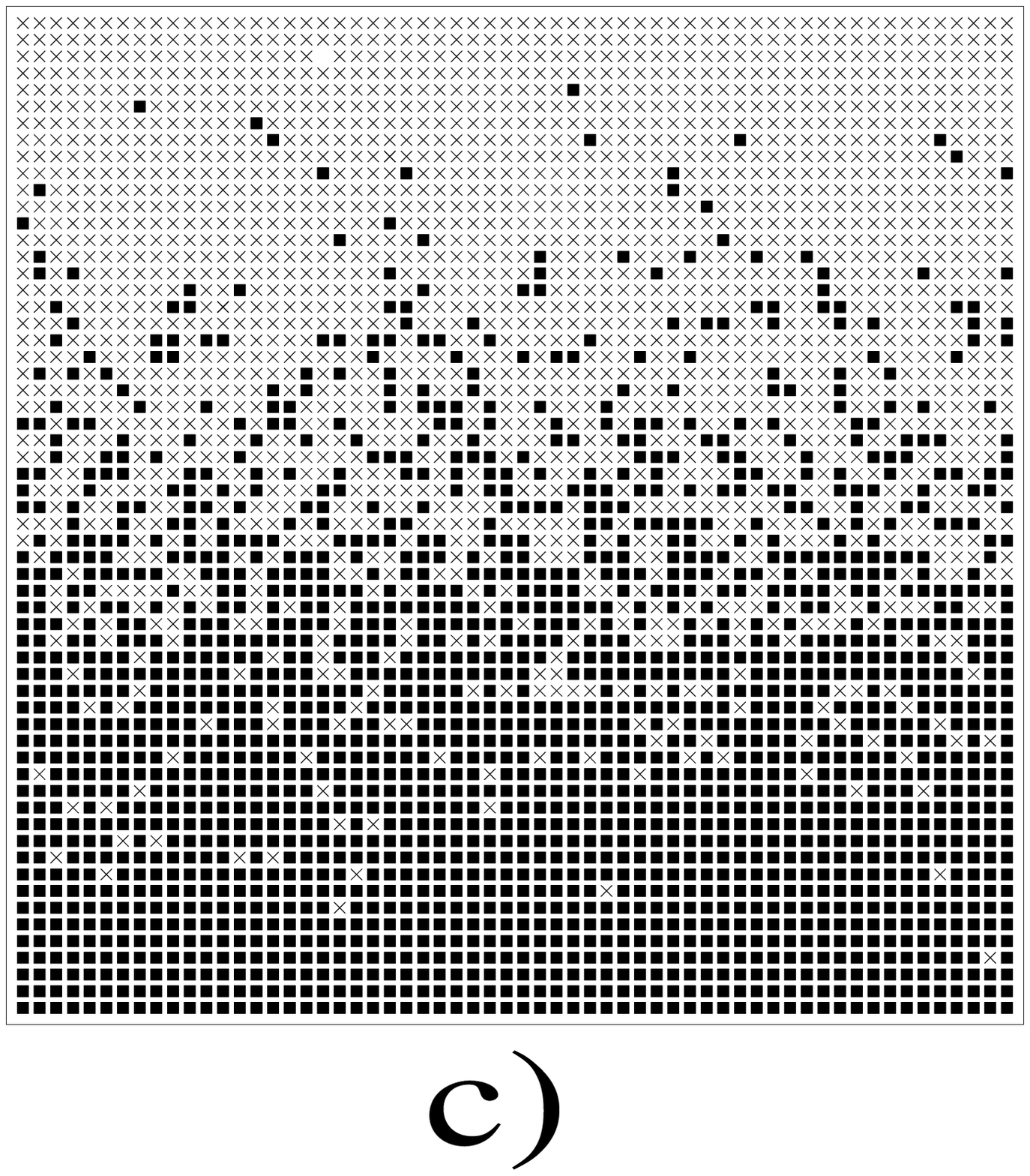}
\end{minipage}\hspace*{-0.5cm}
\begin{minipage}{1.55 in}  
\epsfxsize=1.55 in \epsfbox{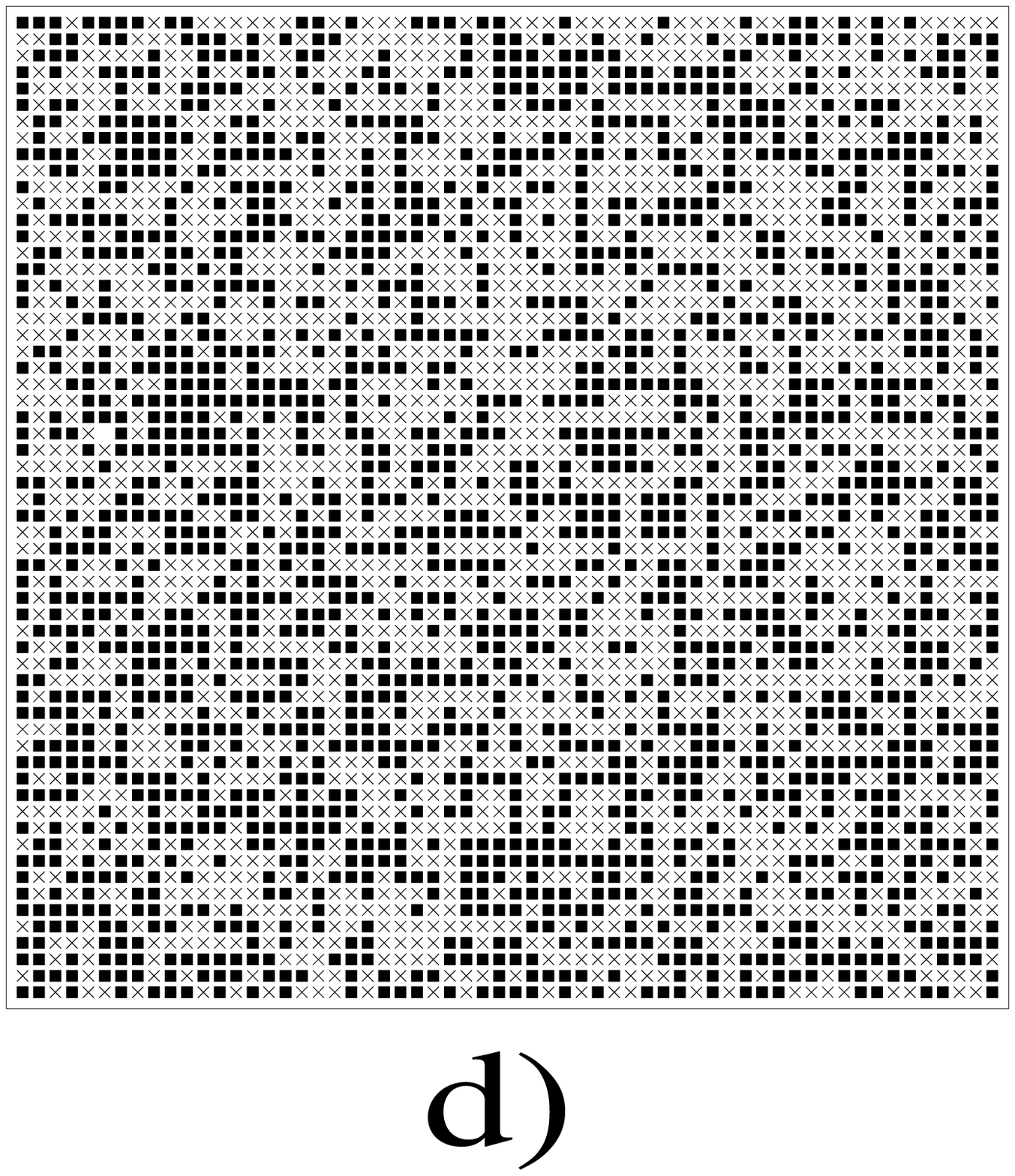}
\end{minipage}
\vspace*{-0.35cm} 
\caption{An initially smooth interface (a)
 gradually becomes rough (b) 
($2\cdot 10^3$ steps), and destroyed after 
(c) (after $5\cdot 10^5$ steps).
(d) represents the completely mixed,
 equilibrium state (more than $2\cdot
10^7$ steps). Here, $L=60$.}
\end{figure} 
 
Our model is defined on a two-dimensional
 square lattice of $L^{2}$ sites, 
with $L$ ranging from $10$ to $100$. 
Each lattice site can be occupied 
either by a white particle, a black 
particle or a hole (vacancy), so that 
multiple occupancy is forbidden. The 
number of vacancies is denoted by $M$. 
For simplicity, we restrict our 
attention only to the case of $L^{2}/2$ 
black particles and $L^{2}/2-M$ 
white particles. Since we wish to describe 
bulk disordering rather than the 
``evaporation'' of the two bulk phases into 
an empty region, we will keep the 
number of holes small, i.e., $M/L^{2}<<1$. 
Initially, the upper (lower) half 
of the system is occupied by white (black) 
particles only, and the vacancies 
are randomly distributed along the flat 
interface. At each Monte Carlo step
 (MCS), a hole is chosen at random and 
changes place, with equal probability,
 with one of its four nearest 
neighbors. All particle-particle 
exchanges are forbidden. The boundary 
conditions at the right and left 
edge are periodic; for the top and bottom 
edges, we have simulated both 
brickwall and periodic boundary conditions, 
without detecting any significant 
differences in the scaling form and 
exponents.

We are interested in the dynamics of
 the interface destruction and bulk 
disordering caused by these vacancies. 
Monitoring the evolution of a typical 
configuration in a 60x60 lattice, shown
 in Fig.1, we observe that the 
interface is gradually destroyed, via
 increasing fragmentation, until it 
ceases to be identifiable. As a 
quantitative measure for the growing 
disorder, we consider the total 
``surface area'' of the black regions, i.e., 
the total {\em number} of black-white 
and black-hole bonds, after $t$ MCS. 
In analogy with the Ising model, we
 will refer to these pairs as ``broken 
bonds''. Averaging this quantity over
 many configurations ($200$ for $L\leq 
40$ and $20$ otherwise) we obtain the 
``disorder parameter'',{\em \ }${\cal A%
}(L,M,t)$, which depends on system 
size $L$, vacancy number $M$, and `time' $%
t$. Since $M/L^{2}\ll 1$, the dominant
 contribution to this quantity is just 
the average number of black-white 
bonds. Fig. 2a shows ${\cal A}(40,1,t)$, 
i.e., the average surface area in a
 $40\times 40$ system containing a single 
vacancy. One clearly distinguishes 
three regimes, drawn schematically in the 
inset: an {\em early }regime (I), 
the {\em intermediate}, or {\em scaling}, 
regime (II), and finally a {\em late } or 
{\em \ saturation }regime (III) in 
which the system has effectively reached 
the steady state. Physically, the 
three regimes are easily interpreted: 
Tracking the position of the vacancy, 
we observe that, for early times, it 
is confined to a small (yet growing) 
region centered on its starting point,
 far from the boundaries of the 
system. Thus, the early regime is 
strongly dependent on the dimensionality 
of the system (see \cite{tz}). 

\begin{figure}[htbp]
\vspace*{-0.6cm}
\begin{minipage}{3 in}\epsfxsize=2.8 in \epsfbox{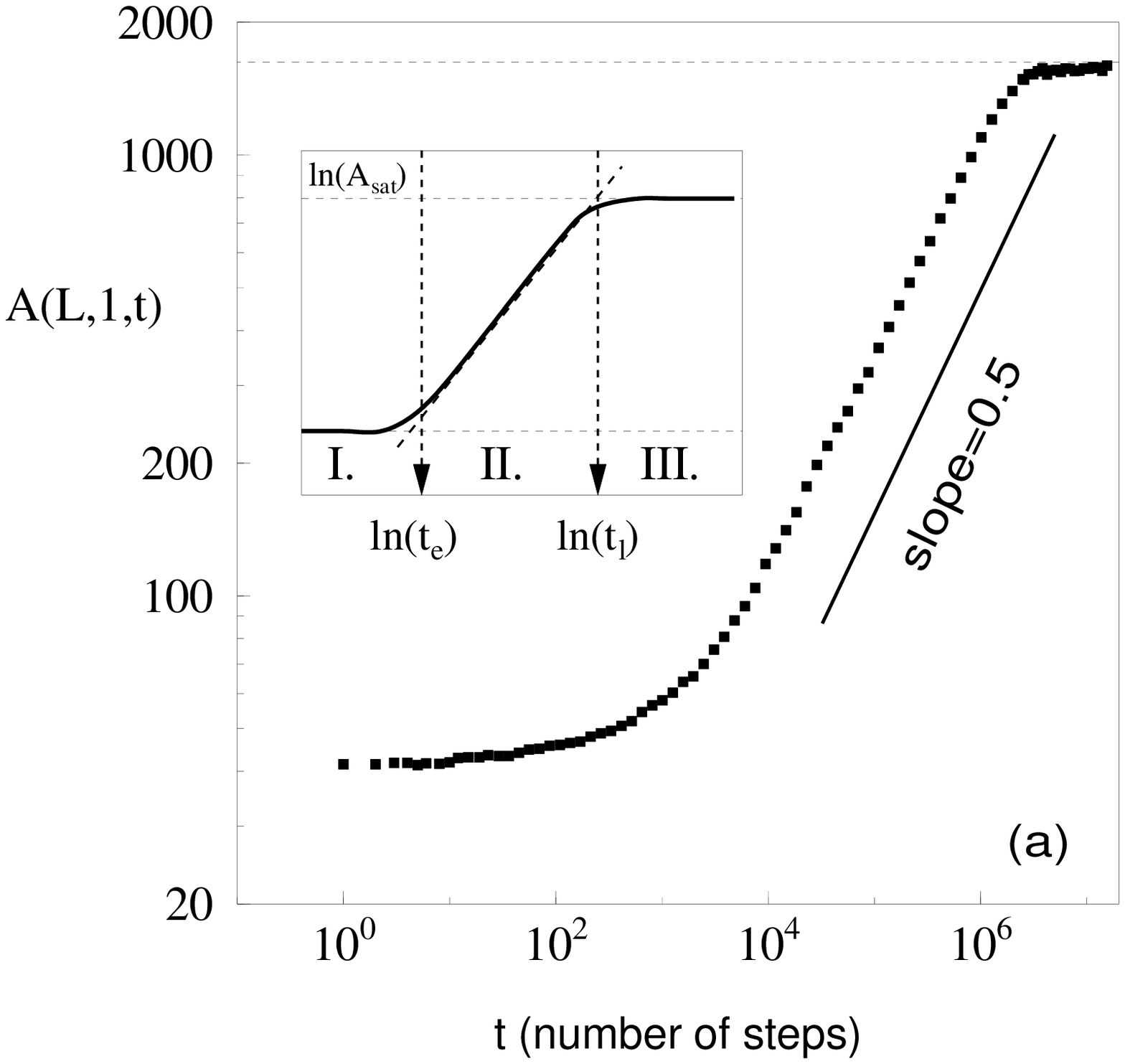}
\end{minipage}
\hspace*{-0.25cm}
\begin{minipage}{3 in}\epsfxsize=2.8 in \epsfbox{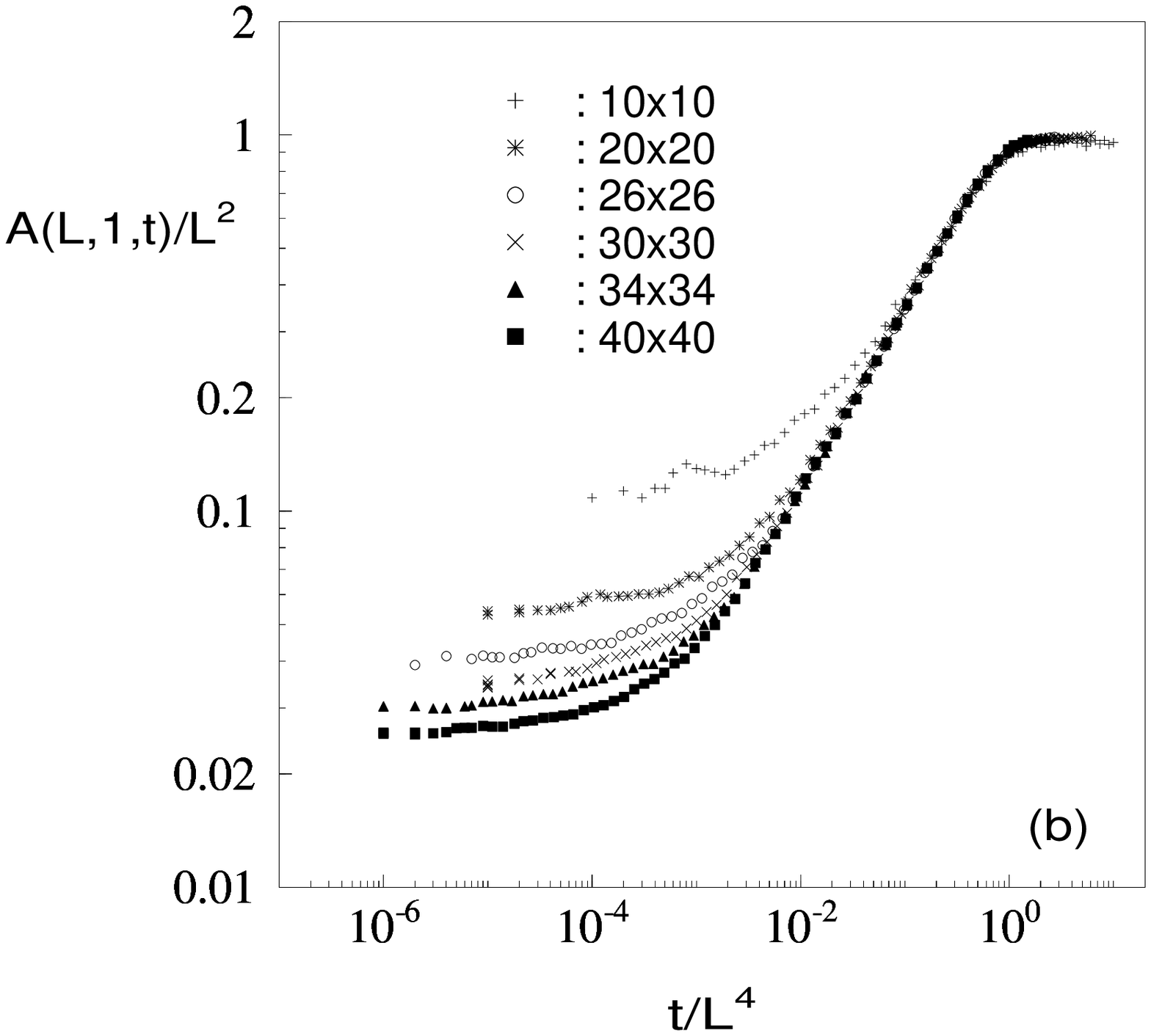} 
\end{minipage}
\vspace*{-1cm}
\caption{ (a) the total number of 
broken bonds, ${\cal A}(L,M,t)$ vs time $t$
(i.e., the number of steps) for 
the single vacancy case $M=1$ and system
size $L=40$. (b) scaling plot of 
${\cal A}(L,1,t)/L^2$ vs $t/L^4$, for
system sizes ranging from $L=10$ to $L=40$.}
\end{figure}

In contrast, for intermediate and late times, 
the vacancy is found with essentially
 equal probability anywhere in the 
system. The difference between 
intermediate and late stages lies in the 
degree of disorder among the black and white particles. 
 
Our key observation is that, independent
 of dimension, Regimes II and III 
exhibit {\em dynamic scaling}. To 
characterize this behavior, we define a 
set of exponents: First, the saturation 
value of ${\cal A}$ scales with 
system size according to $\lim_{t
\rightarrow \infty }{\cal A}(L,1,t)\equiv  
{\cal A}_{sat}(L,1)\sim L^{\alpha }$.
 Second, in the intermediate regime, $%
{\cal A}(L,1,t)$ grows as ${\cal A}
(L,1,t)\sim L^{\sigma }t^{\beta }$. The $L 
$-dependent factor is required here 
to absorb a nontrivial shift along the $%
\ln t$ axis in dimensions other than
 two in order to collapse curves 
associated with different system 
sizes. Finally, the two crossover times 
(``early'' and ``late'') scale as 
$t_{e}\sim L^{z_{e}}$, and $t_{l}\sim 
L^{z_{l}}$. In $d=2$, the data give
 $\alpha =2\pm 0.1$, $\beta =0.5\pm 
0.06$, $z_{e}=2\pm 0.2$, $z_{l}=4\pm 0.2$, 
$\sigma =0\pm 0.1$. 
Since $\sigma$ is effectively zero,
 excellent data collapse is obtained by 
plotting ${\cal A}(L,1,t)/L^{\alpha }$ 
versus $t/L^{z_{l}}$, shown in Fig. 
2b for a range of system sizes. Thus, 
the intermediate-to-late time 
crossover can be summarized by the 
usual Family-Vicsek \cite{fv} scaling 
form  
\begin{equation} 
{\cal A}(L,1,t)\sim L^{\alpha }f
\left( \frac{t}{L^{z_{l}}}\right)  
\label{scalrel1} 
\end{equation} 
with a scaling function $f$, satisfying 
$f(x)\simeq const$ for $x\gg 1$, and  
$f(x)\sim x^{\beta }$ for $x\ll 1$ (but 
large enough to fall within Regime 
II). The consistency condition ${\cal A}
(L,1,t_{l})\simeq {\cal A}_{sat}(L,1) 
$ yields the scaling law  
\begin{equation} 
\sigma +z_{l}\beta =\alpha ,  \label{scalaw1} 
\end{equation} 
which is manifestly satisfied by the 
measured exponents. The same indices 
are observed for several vacancies, provided
 their number remains {\em %
constant} as the system size $L$ is varied. 
We stress that, due to the 
presence of the novel exponent $\sigma $, 
our scaling law (\ref{scalaw1}) is  
{\em distinct} from the familiar $z\beta 
=\alpha $ which controls surface 
growth in, e.g., the Edwards-Wilkinson 
\cite{EW} or KPZ \cite{KPZ} models. 
 
In a real system it is to be expected that 
the number of vacancies itself 
depends on the system size. For generality,
 we allow $M\sim L^\gamma $, where  
$\gamma \in [0,d]$ will be called the
{\em vacancy number exponent}.
 The data reported above then 
correspond to $\gamma =0$ (and $d=2$). 
Another natural case is $\gamma = d-1$,
which could arrise if we imagine welding
two different solids along a common surface
with vacancies uniformly distributed thereupon.
If the two surfaces were of equal fractal dimensions,
a uniform distribution of vacancies on the interface
would lead to a non-integer $\gamma$.
In this study we have examined the case $\gamma =1$, 
i.e., the vacancy number being extensive in 
the initial interfacial length, $M\propto L$.
Here, the data collapse only 
if {\em (i) }curves corresponding 
to different system sizes are shifted by 
an $M$-dependent factor along the $%
\ln t$-axis, and {\em (ii) }the late 
crossover exponent takes on the new 
value $z_l=3\pm 0.2$, while $\alpha $ and
 $\beta $ remain unchanged! This 
behavior suggests that the exponent $\sigma $
 acquires an additional 
significance, depending {\em non-trivially}
 on $\gamma $. For $\gamma =1$, 
the data are consistent with $\sigma =1/2$, 
so that the scaling form (\ref 
{scalrel1}) still holds and the scaling law 
(\ref{scalaw1}) is satisfied 
with the modified exponent $z_l$. 
 
In order to understand the role of $\gamma $, 
we now turn to an analytic 
description of the disordering process, in 
general dimension $d$. For 
simplicity, we choose fully periodic boundary
 conditions on a hypercube of 
side length $L$. The initial value, ${\cal A}
(L,M,t=0)=2L^{d-1}$, reflects 
two flat interfaces between black and white 
particles, with vacancies 
replacing $M$ of the white particles at one 
of the interfaces. Similarly, we 
have ${\cal A}_{sat}(L,M)=\frac d2L^{2d}/(L^d-1)$ 
in the saturation regime, 
since the final steady state is completely 
random. Thus, we can read off the  
{\em exact} result $\alpha =d$. 
 
The remainder of our analysis will be a 
mean-field theory, based on a set of 
equations of motion for the coarse-grained 
local hole and particle densities  
\cite{hsz,long} which are most conveniently
 discussed in continuous time and 
space, in general dimension $d$. A 
well-defined continuum limit is obtained 
by letting the lattice constant $a$ vanish
 and identifying the microscopic 
time scale with $\tau \equiv a^2/2d$. In 
this limit, physical distances such 
as the system size $L$ remain fixed 
while the number of sites approaches 
infinity. For convenience, we define $V\equiv L^d$. 
 
First, we note that each of the vacancies 
performs a simple random walk. As 
a result, the evolution equation for 
$\phi ({\bf x},t),$ the density of 
vacancies, at position ${\bf x}\equiv 
(x_{1},...,x_{d})$ and time $t$ is 
just a diffusion equation, $\partial _{t}
\phi ({\bf x},t)=\nabla ^{2}\phi (%
{\bf x},t)$ , subject to fully periodic
 boundary conditions on $\phi ({\bf x}%
,t)$ and some initial condition. Choosing 
the normalization $\int_{V}\phi (%
{\bf x},t)=M$, we see that, for our 
simulations, $\phi ({\bf x},0)=\frac{M}{%
L^{d-1}}\delta (y-L/2)$ where $y\equiv x_{d}$.
 Note that we still use the 
symbol $t$ here, since the Monte Carlo 
time and its continuum limit differ 
only by the scale factor $\tau $. The 
final state is, of course, trivial: $%
\phi ({\bf x},\infty )=M/V.$ 
 
A similar evolution equation for the 
black particle density, $\psi ({\bf x}%
,t),$ is easily derived from a microscopic
 master equation. Truncating all 
correlations, we find 
 
\begin{equation} 
a^{-d}\partial _{t}\psi ({\bf x},t)=
\phi ({\bf x},t)\nabla ^{2}\psi ({\bf x}%
,t)-\psi ({\bf x},t)\nabla ^{2}\phi ({\bf x},t).  \label{03} 
\end{equation} 
This equation simply tallies the 
local change in black particle density at 
position ${\bf x}$: the first term 
reflects a gain, provided a vacancy is 
initially present and a black particle
 ``diffuses'' in from a neighboring 
site. The second term accounts for a 
loss, due to a black particle jumping 
from ${\bf x}$ to a vacant nearest-neighbor 
site. Similar equations have 
been discussed in the context of 
biased diffusion of two species \cite{hsz}. 
The prefactor $a^{-d}$, appears for
 dimensional reasons. Choosing $\int_{V}\psi 
({\bf x},t)=V/2$, we 
write the initial condition for the 
half-filled system as $\psi ({\bf x},0)
=1-\theta (y-L/2)$. Of course, the 
final state is given by $\psi 
({\bf x},\infty )=1/2.$  
 
The key difference between these 
two equations resides in a separation of 
time scales: the hole density 
relaxes very fast, driven by a diffusion 
coefficient of unity, compared to 
the particle density whose diffusion 
``coefficient'' is $\phi ({\bf x},t)
=O(1/V)$. Performing a simple random 
walk, the vacancies reach the edge of 
the system after a time of order $L^{2} 
$. Thus, we identify the {\em early }
crossover time $t_{e}\propto L^{2}$ and 
read off $z_{e}$ $=2$. For {\em later}
 times, the holes are uniformly 
distributed over the system, and we 
may replace $\phi \ $by its final value,  
$\frac{M}{V}$, so that (\ref{03}) 
reduces to a simple diffusion equation, $%
\partial _{t}\psi =D\nabla ^{2}\psi ,$
where $D\equiv a^{d}M/V$. Its 
solution, subject to the initial and 
fully periodic boundary conditions, is 
easily found:  
\begin{equation} 
\psi ({\bf x},t)=\frac{1}{2}+\frac{2}
{\pi }\sum_{n=1}^{\infty }\frac{%
sin[2\pi (2n-1)y/L]}{2n-1}\;e^{-\epsilon \,t(2n-1)^{2}}\;,  \label{31} 
\end{equation} 
where $\epsilon \equiv 4\pi ^{2}D/L^{2}$. 
 
To make contact with the ``disorder 
parameter'', ${\cal A}$, measured in the 
simulations, we could define an operator 
on the lattice configurations for 
the total number of broken bonds, make a 
meanfield approximation for pair 
correlations, and take the continuum limit
 \cite{long}. Here, let us present 
a short cut, based on the analogy with 
the Ising model and its associated 
coarsed grained version. The local
 magnetization density, $\Phi ({\bf x})$, 
of the former clearly maps into 
$\psi ({\bf x},t)-\frac 12$ for our case. 
Meanwhile, the local energy density 
of the Ising model is given precisely by 
the broken bonds. In the standard 
literature \cite{Amit}, the total energy 
of the Ising model is written as 
$K_1-K_2\int_V\Phi ^2$, where the constants  
$K_i$ may be fixed by the scale and 
the ``zero'' of the energy. Since our 
interest is also the total number 
of broken bonds, we may write ${\cal A}%
\propto KV-\int_V\left[ \psi ({\bf x},t)
-\frac 12\right] ^2$. Note that this 
form expresses the extensivity of 
${\cal A}$. Of course, since the total 
number of broken bonds at $t=0$ is 
only $O(L^{d-1})$, we set the initial 
value of ${\cal A}$ to be $0$. Demanding 
the final value be $dV/2$, we have  
\begin{equation} 
{\cal A}(L,M,t)=\frac d2V-2d\int_V\left[ 
\psi ({\bf x},t)-\frac 12\right] 
^2\quad .  \label{contlim} 
\end{equation} 
 
Using (\ref{31}), we find the time 
evolution of this quantity:  
\begin{equation} 
{\cal A}(L,M,t)=\frac d2V\left[ 
1-S(2\epsilon t)\right] \;,  \label{contlim1} 
\end{equation} 
where $S(\xi )\equiv \frac 8{\pi ^2}
\sum_1^\infty e^{-\xi (2n-1)^2}/(2n-1)^2$%
. Since $S(0)=1$ and $S(\infty )=0$, 
we verify that ${\cal A}$ does take on 
the correct initial and final values. 
To capture the time dependence in the 
intermediate regime (II), i.e., $\epsilon t
\ll 1$, we reexpress the infinite 
sum via a Poisson resummation \cite{Jones}.
 Introducing $u_m\equiv \pi m/2%
\sqrt{2\epsilon t}$, we find 
 
\begin{equation} 
{\cal A}(L,M,t)\simeq \frac{2d}{\pi ^{3/2}}
V\sqrt{2\epsilon t}\left\{ 
1+\sum_{m=1}^\infty (-1)^m\left[ e^{-u_m^2}-
u_m\Gamma \left( \frac 12%
,u_m^2\right) \right] \right\} \;  \label{contlim3} 
\end{equation} 
where $\Gamma (\bullet ,\bullet )$ 
denotes the incomplete Gamma function. In 
this form, the sum over $m$ is suppressed 
for small $\epsilon t$. Thus, $%
{\cal A}(L,M,t)\propto V\sqrt{2\epsilon t}
\propto L^d\sqrt{Mt/L^{2+d}}$, 
yielding the remaining exponents, namely 
$\beta =\frac 12$ independent of 
dimension, and $\sigma =\frac 12(d+\gamma -2)$.
 The late crossover time, 
naturally defined by $\epsilon t_l=1$, 
scales as $t_l\sim L^{2+d-\gamma }$ 
whence one obtains $z_l=2+d-\gamma $. In 
two dimensions our theory predicts $%
z_l=4$ and $\sigma =0$ for $\gamma =0$, 
while $z_l=3$ and $\sigma =\frac 12$ 
for $\gamma =1$ , in complete agreement 
with the Monte Carlo data. 
 
To summarize, we have analyzed the 
disordering process of an initially phase 
segregated system, driven by highly mobile 
Brownian vacancies distributed 
according to the exponent $\gamma $. The 
late stages of the evolution exhibit 
dynamic scaling. A set of exponents 
$\left\{ z_{e},z_{l},  
\alpha, \sigma; \beta \right\} $ can be defined, 
characterizing, respectively, the system 
size dependence of two crossover 
times, the final saturation value of the 
number of broken bonds, ${\cal A}%
(L,M,t)$, and its amplitude in the 
intermediate regime. The temporal growth 
of ${\cal A}$ during the latter regime 
is captured by the exponent $\beta $. 
All indices can be computed analytically,
 in excellent agreement with the 
data. Our key result is that the typical 
time scale $t_{l}\sim L^{z_{l}}$, 
which controls the crossover between 
increasing disorder and saturation, is 
set by $z_{l}=2+d-\gamma $, and thus 
depends explicitly on both the space 
and fractal dimensionalities, $d$ and 
$\gamma $. Measurements of $z_{l}$ can 
therefore provide information about the
 vacancy distribution in a sample. In 
the most familiar case, standard vacancy
 diffusion in solids, the number of 
vacancies is extensive ($\gamma =d$), so
 that the well-known result $z_{l}=2$ 
is reproduced \cite{ms}. Even though our 
model is very simple, it forms the 
basis for the description of a large variety 
of related problems. Work is in 
progress to analyze the effect of external 
driving forces, interparticle 
interactions and vacancy-induced 
catalytic reactions.  
 
\stars
 
Illuminating discussions with 
D. Farkas, C. Laberge and S. Sandow are 
gratefully acknowledged. This 
research is supported in part by the US 
National Science Foundation through 
the Division of Materials Research and 
the Hungarian Science Foundation 
under grant numbers OTKA F17166 and T17493.

\end{document}